\documentclass[conference,a4paper,twocolumn]{IEEEtran}
\def\IEEEkeywordsname{Keywords }
\ifCLASSINFOpdf
\else
\fi
\usepackage{url}
\usepackage{epstopdf}
\usepackage{bm}
  \usepackage{tipa}
  \usepackage[pdftex]{graphicx}
	\usepackage{epsfig}
\usepackage{tabularx}
\usepackage{graphicx}
\usepackage{multirow}
\usepackage[export]{adjustbox}
\usepackage{float}
\usepackage{algpseudocode,algorithm,algorithmicx}
\usepackage{longtable}
\usepackage{amsmath,amssymb,amsfonts}
\usepackage{epstopdf} %converting to PDF
 
 \usepackage{subcaption}
 \usepackage{listings}
\usepackage{color}
\definecolor{babyblueeyes}{rgb}{0.63, 0.79, 0.98}
\definecolor{dkgreen}{rgb}{0,0.6,0}
\definecolor{gray}{rgb}{0.5,0.5,0.5}
\definecolor{mauve}{rgb}{0.58,0,0.82}
\definecolor{bluekeywords}{rgb}{0,0,1}
\definecolor{greencomments}{rgb}{0,0.5,0}
\definecolor{redstrings}{rgb}{0.64,0.08,0.08}
\definecolor{xmlcomments}{rgb}{0.5,0.5,0.5}
\definecolor{types}{rgb}{0.17,0.57,0.68}
\definecolor{cyan}{rgb}{0.0,0.6,0.6}
\definecolor{purple}{rgb}{0.44,0.16,0.39}

\usepackage{algorithm,algpseudocode,algorithmicx,tabularx}
\usepackage{setspace}
\usepackage{verbatim}
\usepackage{amssymb}
\usepackage{amsmath}
\usepackage[justification=centering]{caption}
\usepackage{cleveref}

\crefformat{section}{\S#2#1#3} % see manual of cleveref, section 8.2.1
\crefformat{subsection}{\S#2#1#3}
\crefformat{subsubsection}{\S#2#1#3}
\makeatletter
\newcommand{\multiline}[1]{%
  \begin{tabularx}{\dimexpr\linewidth-\ALG@thistlm}[t]{@{}X@{}}
    #1
  \end{tabularx}
}
\begin{document}
\title{MQTTg: An Android Implementation}

\author{
\IEEEauthorblockN{
Andrew Fisher$^{{1},{2}}$, Gautam Srivastava$^{{1},{3}}$, and Robert Bryce$^{2}$
}\\

\IEEEauthorblockA{
$^{1}$Department of Mathematics and Computer Science, Brandon University, Brandon, Canada\\
$^{2}$Heartland Software, Inc., Ardmore, Alberta, Canada\\
$^{{3}}$Research Center for Interneural Computing, China Medical University, Taichung, Taiwan, Republic of China\\
}
\thanks{Identify applicable sponsor/s here. If no acknowledgments, delete this line.}}
\markboth{}{}
\pagestyle{empty}%
\maketitle%
\thispagestyle{empty}

\begin{abstract}
The Internet of Things (IoT) age is upon us. As we look to build larger networks with more devices connected to the Internet, the need for lightweight protocols that minimize the use of both energy and computation gain popularity. One such protocol is Message Queue Telemetry Transport (MQTT). Since its introduction in 1999, it has slowly increased in use cases and gained a huge spike in popularity since it was used in the popular messaging application Facebook Messenger. In our previous works, we focused on adding geolocation to MQTT, to help modernize the protocol into the IoT age. In this paper, we build off our previous work on MQTTg and build an IoT Android Application that can pull geolocation information from the Operating System. We then use the geolocation data to create geofences to help further tailor the use cases of MQTTg.
\end{abstract}


\IEEEpeerreviewmaketitle

\section{Introduction}
The Internet of Things (IoT) at its core looks to improve our ability as technologists to implement better data sharing, remote control, and data monitoring~\cite{zanella2014internet}. IoT can achieve this by connecting almost any device to the Internet~\cite{dwivedi2019differential,dwivedi2019decentralized}. One of IoT's gems is the recently popular idea of Smart cities and Smart factories, where entire geographic locations or workplaces have devices connected into a network being able to share information~\cite{perera2014sensing}. For smart cities and other IoT ideas, communication protocols for acquisition of large amounts of data and sharing of that data are very important. One such protocol is MQTT.

Vergara \textit{et al.} in \cite{vergara2013mobile} were one of the first to show the superiority of MQTT implemented on Android devices to decrease energy consumption. Furthermore, we now know from a large collection of use cases that MQTT provides some key advantages like a small energy footprint and very low bandwidth use~\cite{Light2017}. Furthermore, even in comparison to other IoT protocols, MQTT also provides advantages in smartphone use~\cite{de2013comparison}. A favorable picture of MQTT based techniques for data communication problems on IoT networks has been painted thus far in related literature~\cite{larmo2019impact,nasr2019pervasive,syed2018modelling,roy2018application}. Our initial work on MQTT started in~\cite{tsp2018,ICSOFT18} we attempted to add geolocation to an existing MQTT release known as Mosquitto~\cite{Light2017}. However, due to shortcomings with the implementation, we switched to MQTTNet~\cite{MQTTnet} and presented our work very recently in~\cite{srivastava2018green}. In this paper we aim to implement the usability of the MQTTg protocol with added geolocation on mobile based applications in the Android Operating System. 

%\textbf{Paper Organization} 
The rest of the paper is organized as follows. In \cref{mqtt} we give a brief overview of MQTT. We follow this with our main contributions in \cref{cont}. We then present main results in \cref{res} followed by some experimental work in \cref{exp}. We end the paper with some future directions in \cref{fw} and end with some concluding remarks in \cref{conc}.

\subsection{MQTT Protocol}\label{mqtt}
The MQTT protocol is a well known publish/subscribe protocol. MQTT relies on publishers to publish content and subscribers to subscribe to given topics to retrieve all messages relating to a given topic. In most implementations, there is a broker that routes information to where it needs to go. Historically, the MQTT protocol runs over TCP/IP and has a data packet size with low overhead minimum ($\geq2$ bytes) so that consumption of power is kept to a minimum. Although we do not discuss the MQTT protocol in depth here, we recommend interested readers to the MQTT documentation in \cite{MQTTnet,MQTTDoc} and to a survey work like \cite{soni2017survey}.

There are so many options for implementing the MQTT protocol on devices. In the scenario presented here, a common system of MQTT requires two main software components:
\begin{itemize}
\item an MQTT Client to be installed on an Android device. A web platform, which uses Javascript, can use the Client PAHO library of Eclipse~\cite{Paho}.
\item an MQTT Broker serves to handle publish and subscribe data. A Linux platform can use a broker that is available for free such as Mosquitto, HiveMQ, etc. In our implementation, we make use of MQTTNet~\cite{MQTTnet}, which is also Open Source.
\end{itemize}

The advantage of the publish/subscribe system is that the data sender (publisher) and the data receiver (client/subscriber) do not know each other because there is a broker in between. In addition, there is time decoupling which makes publisher and client unable to be connected simultaneously to allow the client to not have delays in receiving messages they subscribe to. 
\subsection{Our Contributions}
\label{cont}

We modify both MQTTnet and Paho by adding geolocation information into specific MQTT packets such that, for example, client location could be tracked by the broker and clients can subscribe based not only by topic but also by their specific geolocation. This work is backwards compatible and our modified brokers and clients work with existing code bases when geolocation is not included. A list of all MQTT packets that use geolocation is given in Table~\ref{packets}. This can lead to the client’s last known location having a comparison to a \textbf{polygon geofence}. One important feature of GPS Tracking software using GPS Tracking devices is geofencing and its ability to help track assets. Geofencing allows users of a Global Positioning System (GPS) Tracking Solution to draw zones (geofences) around places of importance, customer’s sites and secure areas. 

\begin{figure}[!h]
\centering
\includegraphics[scale=0.15]{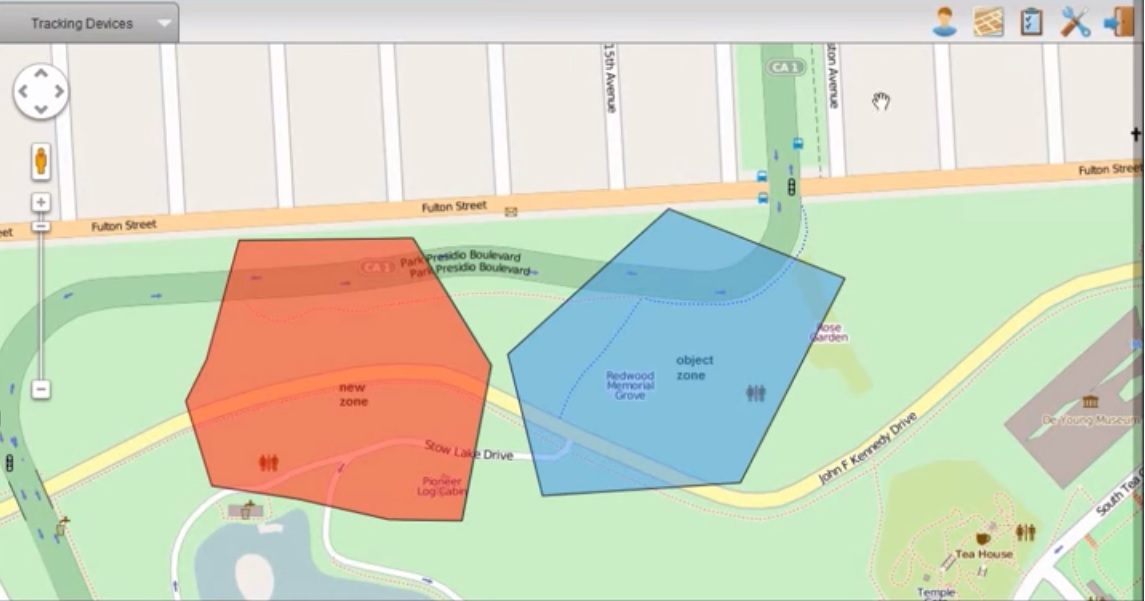}
\caption{Polygon Geofences\cite{wiki}}
\label{fig02}
\end{figure}

In MQTTg, by adding geolocation, information reaching subscribers can be filtered out by the broker to only fall within the subscriber's geofence. We can see an example of a geofence in Figure~\ref{fig02}. As a green IoT example, take a smart city driving condition topic. By prescribing a geofence where driving conditions may not be adequate for a variety of reasons (weather, construction, or an accident for example), specific subscribers on a smart city topic like \texttt{driving conditions} would receive updates based on whether or not their geolocation in real time intersects with a polygon geofence where driving conditions may be abnormal. Other subscribers would receive different messages based on their driving routes throughout the city. We are also motivated by releases such as \texttt{OpenHAB}, an open-source home automation framework~\cite{openhab} and releases like \texttt{OwnTracks}, a private location diary system that allows users on iOS and Android to keep a location diary and share the information with family and friends~\cite{owntracks}, however both releases focus on Payload modifcation not modifying the protocol itself as we do here. Further use cases for MQTTg include: 

\begin{itemize}
\item Field team coordination
\item Search and rescue improvements
\item Advertising notifications to customers within specific ranges
\item Emergency notifications, such as inclement weather or road closures.
\item Taxi cab monitoring and deployment strategies
\end{itemize}

\begin{figure}[ht]
\centering
\includegraphics[scale=0.4]{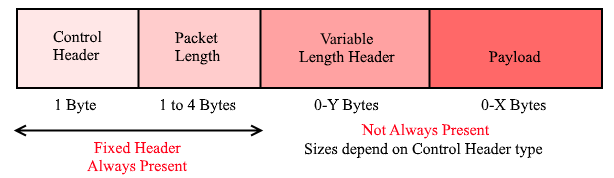}
\caption{MQTT Geolocation Packet}
\label{fig03}
\end{figure}

\begin{figure}[ht]
\centering
\includegraphics[scale=0.35]{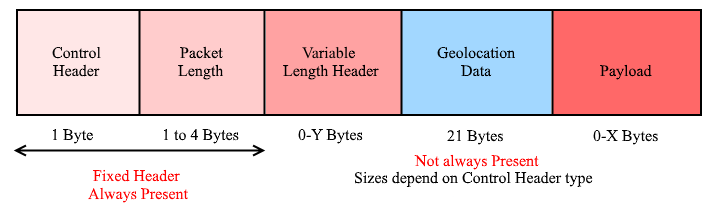}
\caption{MQTT Geolocation Packet}
\label{fig04}
\end{figure}

\section{Results}
\label{res}
The basis of MQTTg is to leverage unused binary bit data within the protocol definition and, optionally, embedding geolocation data between the header and payload as shown in Figure~\ref{fig04}. We can clearly see while comparing Figure~\ref{fig03} to Figure~\ref{fig04} where the changes have been made indicated in \textcolor{babyblueeyes}{blue}. We show details of the $21$ bytes of Geolocation data as shown in Figure~\ref{fig04} in Listing~\ref{struct}.
 
\lstset{language=[Sharp]C}
\begin{lstlisting}[caption={Geolocation Data Layout}\label{struct}]
GeoLocation {
      byte version;
      double latitude, longitude;
      float elevation;
};
\end{lstlisting}
The major change to the packets themselves was the inclusion of the \texttt{Geolocation Flag}. The flag is sent in packets to the broker to notify the broker that they are sending geolocation data in the packet. The packets that are used to send geolocation information are given in Table~\ref{packets}, derived from the original protocol implementation. In Listing~\ref{csh}, we see the updated C\# code for MQTTnet packet deserializer for the \texttt{PUBLISH/PUBLISHG} packet. The \texttt{isGeog} Boolean passed is based on the packet type identified by the calling method. Based on this geolocation flag, we treat the \texttt{PUBLISH/PUBLISHG} packets differently. 
 
\begin{table}[ht]
\centering
\caption{Types of MQTT Packets used for Geolocation}

\begin{tabular}[b]{l@{\hskip 0.15in}l}
\hline\noalign{\smallskip}
\textbf{Packet} & \textbf{Description} \\ \hline\hline
\texttt{PUBLISHG} & Publish message \\
\hline
\texttt{PUBACK} & Publish acknowledgement \\
\hline
\texttt{PUBREC} & Publish received (assured delivery part 1) \\
\hline
\texttt{PUBREL} & Publish received (assured delivery part 2) \\
\hline
\texttt{PUBCOMP} & Publish received (assured delivery part 3) \\
\hline
\texttt{SUBSCRIBE} & client subscribe request \\
\hline
\texttt{UNSUBSCRIBE} & Unsubscribe request \\
\hline
\texttt{PINGREQ} & PING request \\
\hline
\texttt{DISCONNECT} & client is disconnecting \\
\noalign{\smallskip}
\hline
\noalign{\smallskip}
\end{tabular}
\label{packets}
\end{table} 

\lstset{language=[Sharp]C}
\lstset{
 morekeywords={var,ushort},emph={MqttBasePacket,MqttPacketReader,MqttPacketHeader,fixedHeader,MqttGeog,MqttPublishPacket}
}
\begin{lstlisting}[caption={C\# Code from the MQTTnet Packet De-Serializer}\label{csh}]
DeserializePublish
{
            fixedHeader =  mqttPacketHeader;
            qualityOfServiceLevel = fixedHeader.Read(2);

            topic = reader.ReadString();

            if (isGeog)
            {
                GeoLocation.version = reader.ReadByte();
                GeoLocation.latitude = reader.ReadDouble();
                GeoLocation.longitude = reader.ReadDouble();
                GeoLocation.elevation = reader.ReadSingle();
            }

            return packet;
}

\end{lstlisting}
%\captionof{lstlisting}{C\# Code from the MQTTnet Packet De-Serializer}

From the Paho MQTTg implementation, Listing~\ref{java} gives the updated Java implementation for de-serializing MQTT packets which they call \texttt{MqttWireMessage}. For a \texttt{PUBLISH} packet, the Java client is normally setup to determine the topic when creating a new \texttt{MqttPublish} object. For a geolocation packet, it is setup to have the topic before the geolocation data. So, the code is modified to do as such if one is received. The Java client stores the geolocation data in big endian \texttt{IEEE 754} format. The geolocation data is, however, encoded in little endian so the bytes need to be reversed to get the correct output for this data. To be consistent, we are adhering to the \texttt{IEEE} floating point representations everywhere.

\lstset{language=Java}
\lstset{
 morekeywords={var,ushort},emph={MqttBasePacket,MqttPacketReader,MqttPacketHeader,fixedHeader,MqttGeog,MqttPublishPacket, MqttWireMessage, createWireMessage,CountingInputStream,DataInputStream,MqttWireMessage}
}
\begin{lstlisting}[caption={Poha Java Code Packet De-Serializer}\label{java}]
DeserializePublish{
  firstByte = in.readByte();
  type = (first >> 4);
  info = (first &= 0x0f);

  MqttWireMessage result;
 
  MqttGeog GeoLocation = null;
  String topic = null;

  if (type == PUBLISHG) {

   topic = new String(encodedString, "UTF-8");
   
   GeoLocation.version = in.readByte();

   while(i < 8) {
    lat[7 - i] = in.readByte();
    i++;
   }
   GeoLocation.latitude = lat.getDouble();

   while(i < 8) {
    lon[7 - i] = in.readByte();
    i++;
   }
   GeoLocation.longitude =lon.getDouble();
   
   while(i < 4) {
    elev[3 - i] = in.readByte();
    i++;
   }
   GeoLocation.elevation = elev.getFloat();
  }
     
  long remainder = totalToRead - counter.getCounter();
  byte[] data];

  if (remainder > 0) {
   in.readFully(data, 0, data.length);
  }
\end{lstlisting}
%\captionof{lstlisting}{Poha Java Code Packet De-Serializer}

For all packets mentioned in Table~\ref{packets}, with the exception of \texttt{PUBLISH}, the $3$rd bit of the fixed header is unused (reserved) in the original implementation in \cite{Stan1999}, so we can easily use it to indicate the presence of geolocation information. Figure \ref{fig03} shown earlier and Figure \ref{fig04} explain where the location data is in the packet.

The \texttt{PUBLISH} control packet needs a different implementation. Because the $3$rd bit is already allocated for Quality of Service (\texttt{QOS}), and all other packets are also reserved for an existing use, we chose to implement a new control packet type. \texttt{PUBLISHG} (\texttt{=0xF}) is used as the flag type for geolocation data when it is to be sent. There are $16$ available command packet types within the MQTT standard and $0$ through $14$ are used. 

We deem geolocation data as an optional attribute, as not all clients may wish to publish their geolocation data for security reasons. In our approach, geolocation data is not included in the packet payload, since not all packet types support a payload, thus rendering payloads not a viable option--- especially for green IoT. Furthermore, we did not wish the broker to examine the payload of any packet, thus keeping our processing footprint low.

%\subsection{Justification of Overhead}
%
%There are many transport protocols that MQTT packets may go over in size, where 21 bytes may be a concern. However, if you are using TCP/IP as we discuss here, 21 bytes is negligible when considering the bytes required just for the overhead of TCP/IP.  Finally, there is no need for a client to only submit MQTTg (geolocation) packets; there is nothing stopping a client to submit MQTTg packets when it is time to update location, then just use unmodified MQTT packets otherwise. The logic is based on a single bit, and we have not introduced any requirement that all packets must be MQTTg – either from a specific client or from multiple clients.

\subsection{Handling of packets}
Packets that are received without geolocation data are handled via the original MQTTnet and Paho functions respectively.  Packets that are received with geolocation are handled similarly but with a call to a \texttt{last known location} updating method--- which stores the client’s unique $ID$ and the location data into a \texttt{Hashtable} object designed to be compared against the geofence. If and only if they are a subscriber is the packet to be sent with geolocation data.  We have elected to attach geolocation data from all packet types originating from the client to eliminate the need for specific packets carrying only geolocation data, and thus reducing network traffic as well. 
	
\subsection{Geofencing}
Creating the geofence code was a major part of MQTTg. The geofence filtering is only called when a packet is submitted to the broker as these packets are forwarded to subscribing clients. 

Geofence data is presently submitted and cleared by a client to the broker using the MQTT \texttt{SUBSCRIBE} packet so that clients may individually submit geofences of interest.  The broker maintains polygon data for each subscribing client.  Polygons may be \texttt{static} in shape and location or \texttt{dynamic} and move with the last known location of the target.

Since the $3$rd bit of the fixed header is unused, we are able to indicate whether or not the  \texttt{SUBSCRIBE} packet contains \texttt{GeoLocation} data. If the data is present, the next $21$ bytes are read as before to get this information. The next step is to read the topic filter for the packet. In our implementation, we have created two new \texttt{GeoLocation} filters called \texttt{ForwardInsideRadiusTopicFilter} and \texttt{ForwardOutsideRadiusTopicFilter} respectively. With each one, the client can subscribe based on a topic, latitude, longitude, and radius to either receive \texttt{PUBLISHG} packets from within or outside of the area. To determine if a \texttt{GeoLocation} filter is present, the second bit in the \texttt{QoS} byte is used as the identifier. If the bit is present, the next byte read will determine which type of \texttt{GeoLocation} filter was passed and it will then proceed to read the radius, latitude, and longitude to create the filter. Our implementation still allows for a normal topic filter to be used.

\begin{figure}[!htb]
    \centering
    \begin{subfigure}[b]{0.2\textwidth}
        \includegraphics[width=\textwidth]{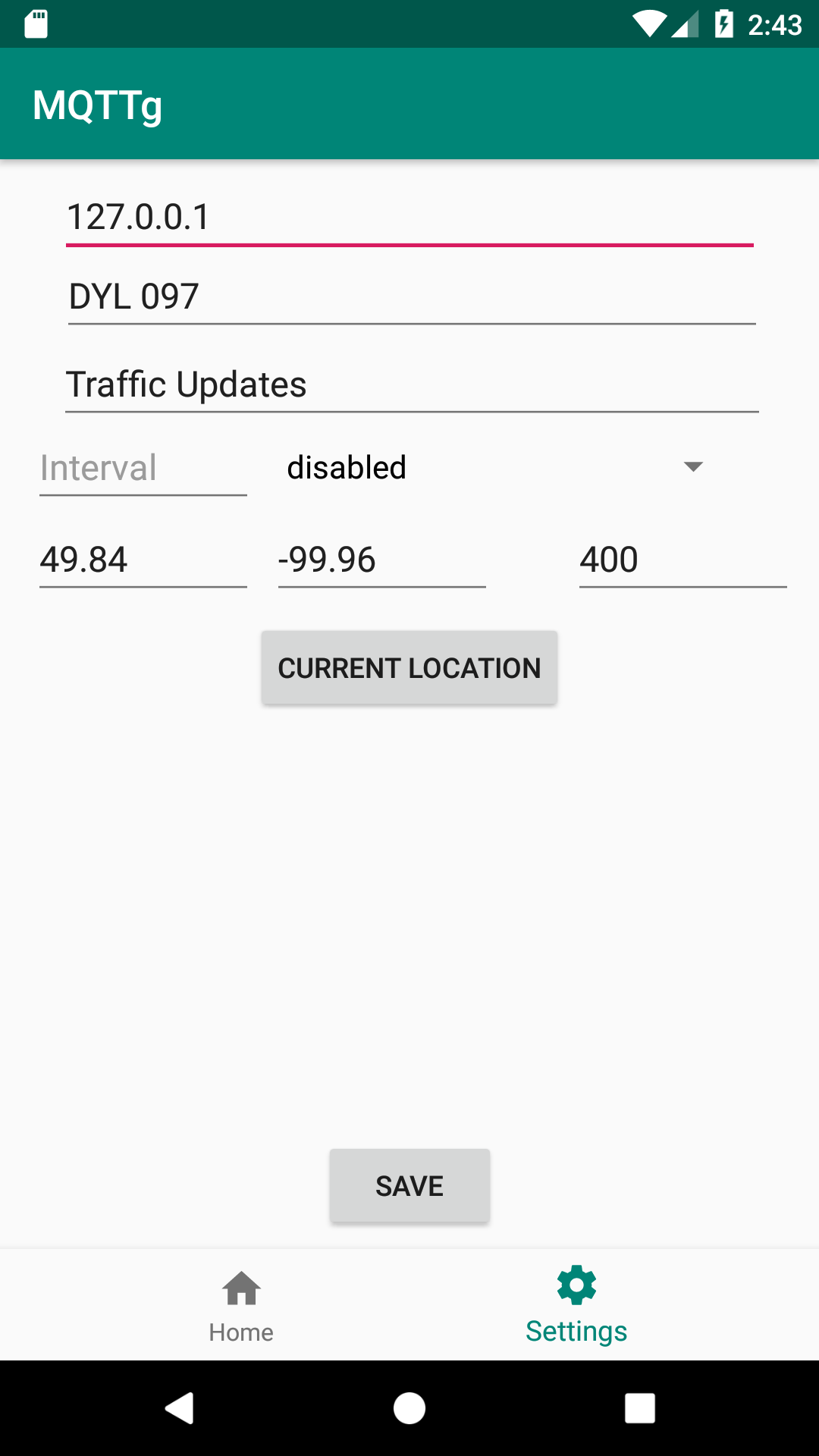}
        \caption{OS Subscriber ID Page}
        \label{os1}
    \end{subfigure}
    ~ %add desired spacing between images, e. g. ~, \quad, \qquad, \hfill etc. 
      %(or a blank line to force the subfigure onto a new line)
    \begin{subfigure}[b]{0.2\textwidth}
        \includegraphics[width=\textwidth]{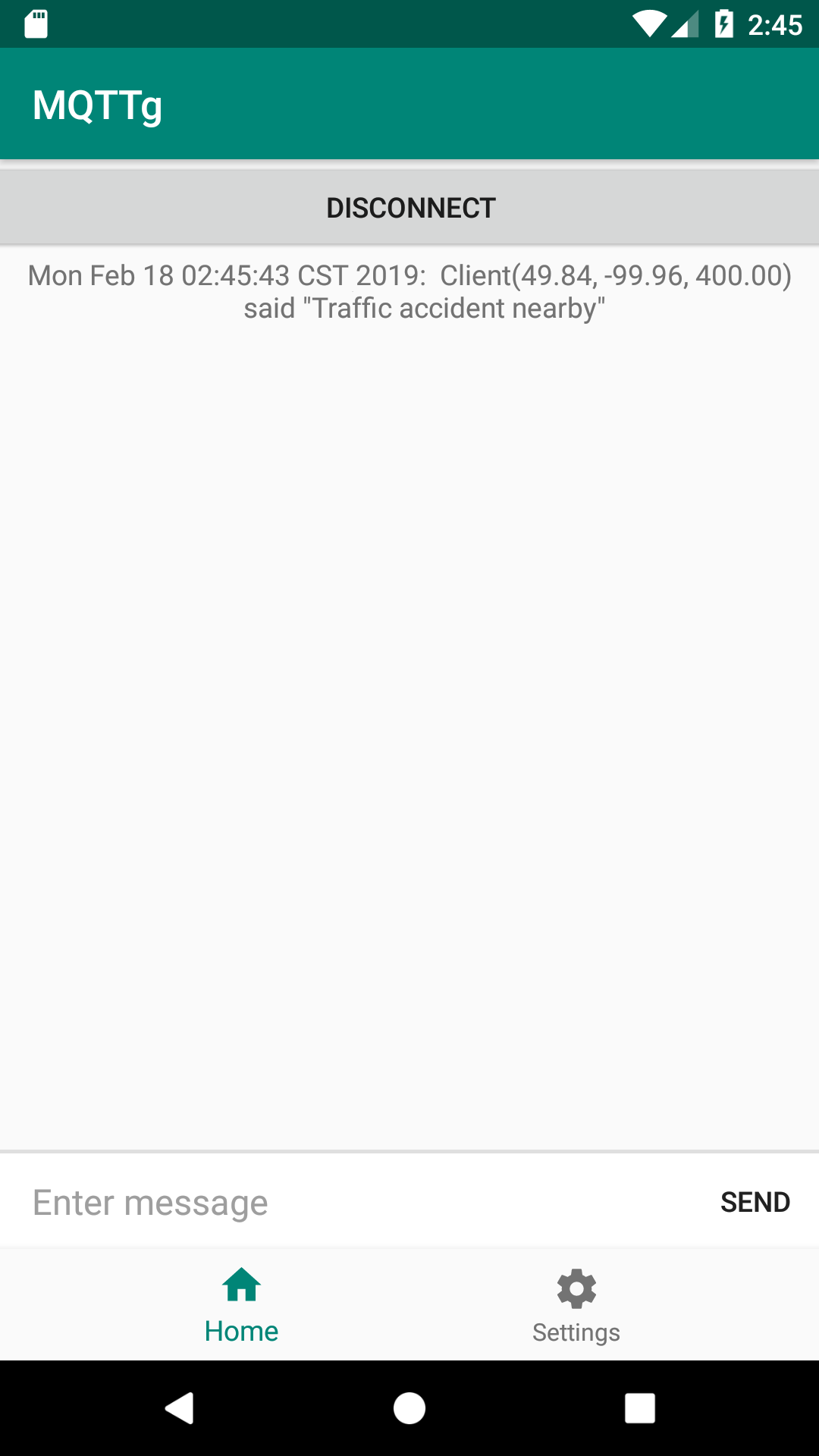}
        \caption{Subscriber Feed Page}
        \label{os2}
    \end{subfigure}
    \caption{Android OS App}\label{android}
\end{figure}

\subsection{Android OS Application}\label{android1}
Figure~\ref{android} provides some snapshots of the current implementation of the Android OS Application for MQTTg. In Figure~\ref{os1}, a subscriber (client) can identify themselves on the network. Pressing the \texttt{Current Location} button will provide the application with the client's current geolocation data pulled from the OS. By not pressing \texttt{Current Location}, the given client acts in original MQTT form lacking any geolocation activity. The topic, say \texttt{Traffic Updates}, will subscribe the client to that topic for future updates, which will show in Figure~\ref{os2}. If an update is provided to the topic by a publishing client, all other clients within a geofence bounded area of the publisher's creation will receive the message. A client can subscribe to as many topics, with or without geolocation, as they choose. In Figure~\ref{os2}, all subscribed topic messages will be shown here. Topics where geolocation are shared will be specific to a given geofence so only matching geolocation data to a given geofence will show. These matches are determined by the broker based on the rules of the geofence from the subscriber. We expect to add separate layouts for a publisher scenario versus a subscriber scenario on the network.
\section{Experimental Results}
\label{exp}
To experiment with the accuracy of geolocation within the MQTTg Android application, we performed a series of experiments. First, a route was mapped within city limits using a well known geolocation user and service, Google Maps~\cite{svennerberg2010beginning}.

\begin{figure}[h!tb]
    \centering
        \includegraphics[width=0.4\textwidth]{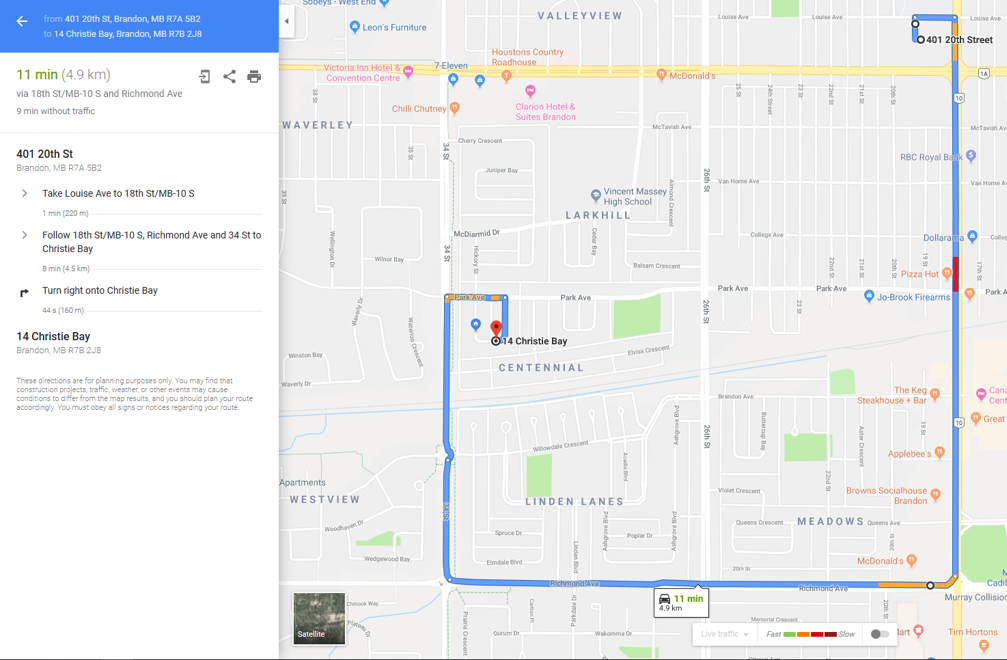}
    \caption{Google Maps Screenshot: Mapping of a Path from Brandon University}
     \label{google}
\end{figure}

\begin{figure}[h!tb]
    \centering
        \includegraphics[width=0.4\textwidth]{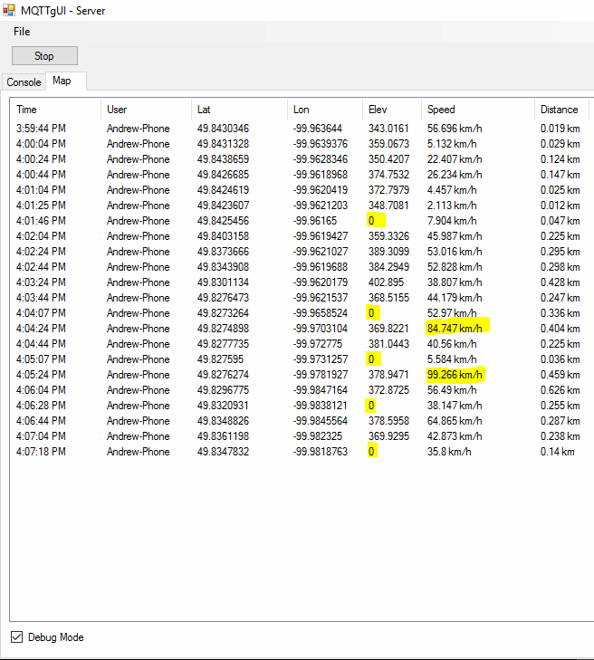}
    \caption{MQTT Broker Screenshot}
     \label{mqttresults}
\end{figure}

In Figure~\ref{google}, we show a Google Maps screenshot of a mapped route from Brandon University to a residential area in Brandon, Manitoba, Canada. The total distance as shown is $4.9$ km. We then, did multiple trials driving the route with an Android Phone, running the MQTT client as described in \cref{android1} with geolocation packets being sent every $30$ seconds sending all relevant geolocation data. Using the Distance measurements calculated by the Broker using geolocation information between packets received, we calculated a total distance traveled being $4.902\pm0.001$ km for all trials. This gives an accuracy rating for the geolocation data of $99.9\%$. Since Google Maps data is only to a precision level of $1$, it may very well be that we achieved a $100\%$ accuracy rating in our trials.

\section{Future Work}
\label{fw}
We are still finishing the final testing of MQTTg for the Android OS. In Figure~\ref{mqttresults}, there were some anomalies both in Elevation and Speed as highlighted. We wish to troubleshoot these issues to see what the cause of this could be.

Applications of the Android client are plentiful but have some key uses in green IoT, natural disaster containment, and safety in this age of mobile devices and smart cities. We can also see some direct applications for visually impaired individuals trying to navigate smart cities \cite{al2015}. Additionally, there is room to make the Android operating system application more visually appealing.

We have yet to deal with both security limitations of MQTT and Quality of Service (\texttt{QoS}) levels and how they will relate to MQTTg. The OASIS standard implemenation of MQTT strongly recommends a MQTT security solution using SSL/TLS \cite{errata2015mqtt}. However, this solution entails additional significant communication and computation overheads for certificate validation checks which may not be feasible in IoT solutions. 

The \texttt{QoS} level is an agreement between the sender of a message and the receiver of a message that defines the guarantee of delivery for a specific message. There are $3$ \texttt{QoS} levels in MQTT that give clients the power to choose a level of service that matches their network reliability and application logic. Therefore, there still needs to be some connection between security, \texttt{QoS} and MQTTg. We are currently working on perhaps applying them together with differing levels of security depending on with \texttt{QoS} in being used. There is room and viability to use the $21$ bytes of information as shown in Figure~\ref{fig04} to help manage the \texttt{QoS} levels and security.

\section{Conclusion}
\label{conc}
In this paper we furthered some initial work on our protocol MQTTg aptly named for the addition of geolocation to the protocol. We were able to create an Android application using MQTTg. We also ran some experimental trials to test the accuracy of MQTTg on the Android OS with strong results. It is important to further this study on MQTTg as their is still both room for improvement as well as baselines needed for performance of the protocol under implementation loads.

\bibliographystyle{IEEEtran}
\bibliography{local_biblio}

\end{document}